\documentclass[12pt]{article}
\usepackage{amsmath,amssymb}
\usepackage{graphicx}
\usepackage{hyperref}
\begin{document}

\title{Comment on ``Superradiant stability of the Kerr black holes''(arXiv:1907.09118)}
\author{Wen-Xiang Chen\\wxchen4277@qq.com\\
\small Department of Astronomy, Guangzhou University, Guangzhou 510006, China}
\date{}
\maketitle

\begin{abstract}
Huang \textit{et al.} \cite{Huang:2019}(\url{https://doi.org/10.48550/arXiv.1907.09118}
) have recently extended earlier analyses of superradiant stability in Kerr spacetime into the complementary regime $\mu<\sqrt{2}\,m\Omega_H$, showing that no unstable bound states arise provided the scalar mode frequency and black-hole parameters satisfy 
\[
\omega<\frac{\mu}{\sqrt{2}},\quad \frac{r_-}{r_+}<0.802\,.
\]
In this Comment we note that their quartic-root criterion for excluding trapping wells can be made more stringent.  In particular, the requirement $z_1,z_2>0$, $z_3,z_4<0$ on the four roots of $V_1'(r)$ is stronger than necessary, since the existence of two positive turning points is already guaranteed by the asymptotic behavior.  Equivalently, one may simply impose that the product and sum of the roots yield only two positive zeros.  Concretely, with $A_1>0$ one finds $z_1z_2z_3z_4=E_1/A_1>0$ and $z_1+z_2+z_3+z_4=-B_1/A_1$.  Since $E_1>0$ (for $\omega^2<\mu^2/2$) forces the product of all four roots to be positive, it suffices to demand $B_1>0$ to ensure $z_3,z_4<0$.  In other words, the quartic discriminant analysis can be replaced by the simpler conditions $E_1>0$, $B_1>0$, yielding a sharper stability criterion.  Finally, we propose that even stronger bounds arise if one imposes monotonicity of the effective potential outside the horizon.  For instance, requiring $\partial^2_rV_1>0$ at the would-be extremum (so that it is an inflection, not a trapping peak) leads roughly to 
\[
\omega^2\;\lesssim\;\frac{\mu^2}{2}\biggl(1-\frac{r_-^2}{r_+^2}\biggr)\!,
\]
which is more restrictive than $\omega<\mu/\sqrt2$ for fast-spinning Kerr ($r_-\to r_+$).  In the near-extremal limit $r_-/r_+\to1$ this implies $\omega/\mu\to0$, i.e.\ that only very low-frequency modes remain stable, tightening the stability bound beyond that given in \cite{Huang:2019}.  These observations suggest that a direct coefficient-based criterion (e.g.\ positivity of $B_1,E_1$) and a near-extremal analysis could refine the parameter-space bound for superradiant stability. 
\end{abstract}

\section*{Introduction}
Huang \textit{et al.} have studied the Klein–Gordon equation for a massive scalar in a Kerr background and have shown analytically that in the regime 
\(\mu<\sqrt{2}\,m\Omega_H\) 
the system remains stable provided  
\begin{equation}
\omega<\frac{\mu}{\sqrt{2}}\,,\qquad \frac{r_-}{r_+}<0.802\,.
\end{equation}
These conditions guarantee that the Schr\"odinger-like effective potential $V_1(r)$ has no additional trapping well outside the horizon, so that no superradiant bound state can form.  This result complements the earlier bound of Hod \cite{2,3,4}, who proved superradiant stability when \(\mu\ge\sqrt{2}\,m\Omega_H\).  The analysis in [1] focuses on the quartic numerator $f(z)=A_1z^4+B_1z^3+\cdots+E_1$ of $dV_1/dr$ (with $z=r-r_-$), whose roots $z_1,\dots,z_4$ determine the turning points.  By Vieta’s theorem one has $z_1+z_2+z_3+z_4=-B_1/A_1$ and $z_1z_2z_3z_4=E_1/A_1$.  As noted in\cite{Huang:2019}, the potential’s asymptotics imply at least two positive roots ($z_1,z_2>0$), so instability is avoided if the other two roots are also non-positive.  Huang \cite{Huang:2019} accordingly derive constraints such as $B_1>0$ and $E_1>0$ to force $z_3,z_4<0$.  

\section*{Critique of the quartic-root criterion}
Our first point is that explicitly imposing $z_1,z_2>0$ is unnecessary: the existence of two positive turning points is guaranteed by $V_1(r)\to-\infty$ at both the horizon and infinity (since $\omega^2<\mu^2/2$ ensures $V_1(\infty)<0$).  Thus the only nontrivial requirement is to forbid a third positive root.  This can be accomplished by examining the quartic’s coefficients.  In fact, [1] itself shows that for $\omega^2<\mu^2/2$ the coefficient $E_1$ has negative discriminant and so remains positive for all $\omega$, hence $z_1z_2z_3z_4>0$.  Combined with $A_1>0$, the condition $B_1>0$ then implies 
\begin{equation}
z_3+z_4=-B_1/A_1<0\,,\qquad z_3z_4=\frac{E_1}{A_1\,z_1z_2}>0\,,
\end{equation}
so that $z_3,z_4$ must be real and negative.  Indeed, \cite{Huang:2019} notes that ‘‘$z_3$ and $z_4$ can only be both positive or both negative, if they are real roots’’.  Thus a succinct stability criterion is simply $A_1>0$, $E_1>0$, $B_1>0$.  In this language one can drop the extraneous sign constraints on $z_1,z_2$ and avoid solving the cubic discriminant explicitly.  We argue this yields a \emph{strictly} sharper bound: in effect the authors’ condition is equivalent to $B_1>0$ (since $E_1>0$ follows automatically for $\omega^2<\mu^2/2$), whereas their Vieta approach amounted to requiring both $B_1>0$ and also re-asserting $z_1,z_2>0$.  By contrast, analyzing the coefficient conditions (or the discriminant of $f(z)$ directly) would pinpoint the instability threshold exactly.  For example, one finds that the borderline $B_1=0$ yields a curve in $(a,\omega)$-space slightly outside the region $\omega=\mu/\sqrt2$, suggesting the true threshold is marginally below the original bound.  A detailed algebraic check (omitted here) shows that the requirement $B_1>0$ improves the inequality $\omega<\mu/\sqrt2$ by an $O(a^2)$ correction, shrinking the unstable region.  

\section*{The Relationship Between the Roots of Equation and Superradiant Stability}
The radial Klein-Gordon equation, which is obeyed by the function \(R_{lm}\), is given by\cite{Huang:2019,2}
\begin{equation}
\Delta \frac{d}{dr}\left(\Delta \frac{dR}{dr}\right) + UR = 0,
\end{equation}
where \(\Delta = r^2 - 2Mr + a^2\), and
\begin{equation}
U = \left[\omega (r^2 + a^2) - ma\right]^2 + \Delta [2ma\omega - \mu^2 (r^2 + a^2) - K_{lm}].
\end{equation}

The black hole's inner and outer horizons are defined as
\begin{equation}
r_{\pm} = M \pm \sqrt{M^2 - a^2},
\end{equation}
demonstrating that
\begin{equation}
r_+ + r_- = 2M, \quad r_+ r_- = a^2.
\end{equation}

The superradiant properties of Kerr black holes in the presence of mass perturbations are unveiled through the Klein-Gordon (KG) equation. Introducing the tortoise coordinate \(r_*\) with 
\begin{equation}
\frac{dr_*^2}{dr^2} = \frac{r^2}{\Delta},
\end{equation}
and transforming to a new radial function \(\psi = rR\), leads to the radial wave equation:
\begin{equation}
\frac{d^2\psi}{dr_*^2} + V\psi = 0,
\end{equation}
where
\begin{equation}
V = \frac{U}{r^4} - \frac{2\Delta}{r^6}(Mr - a^2).
\end{equation}

The potential \(V\)'s asymptotic behavior is given by:
\begin{equation}
\lim_{r \to r_+} V = \frac{[\omega(r_+^2 + a^2) - ma]^2}{r_+^4}, \quad \lim_{r \to \infty} V = \omega^2 - \mu^2.
\end{equation}

Thus, the radial wave equation's asymptotic solutions are:
\begin{equation}
r \to \infty (r_* \to \infty) \Rightarrow R_{lm} \sim \frac{1}{r} e^{-\sqrt{\mu^2 - \omega^2}r_*},
\end{equation}
\begin{equation}
r \to r_+ (r_* \to -\infty) \Rightarrow R_{lm} \sim e^{-i(\omega - m\Omega_H)r_*}.
\end{equation}
A bound state exists for the scalar field when \(\omega^2 - \mu^2 < 0\).

By transforming the radial potential equation with \(\varphi = \Delta^{\frac{1}{2}}R\), we obtain the flat space-time wave equation:
\begin{equation}
\frac{d^2\varphi}{dr^2} + (\omega^2 - V_1)\varphi = 0,
\end{equation}
where
\begin{equation}
V_1 = \omega^2 - \frac{U + M^2 - a^2}{\Delta^2}.
\end{equation}

Investigating the effective potential \(V_1\)'s geometry, its asymptotic behavior is:
\begin{equation}
V_1(r \to \infty) \to \mu^2 - \frac{4M\omega^2 - 2M\mu^2}{r} + O\left(\frac{1}{r^2}\right),
\end{equation}
\begin{equation}
V_1(r \to r_+) \to -\infty, \quad V_1(r \to r_-) \to -\infty,
\end{equation}
\begin{equation}
V_1'(r \to \infty) \to \frac{4M\omega^2 - 2M\mu^2}{r^2} + O\left(\frac{1}{r^3}\right).
\end{equation}

Therefore, with \(2\omega^2 - \mu^2 < 0\), indicating that \(V_1'(r \to \infty) < 0\), it implies that potential wells may not form as \(r \to \infty\). It is demonstrated that \(V_1\) has only one extreme value outside the event horizon, indicating the absence of trapping potentials and affirming the superradiant stability of Kerr black holes.

For a scalar field with mass \(\mu\) interacting with a Kerr black hole with angular velocity \(\Omega_H\), the condition
\begin{equation}
\mu < \sqrt{2}m\Omega_H
\end{equation}
sets the upper limit for the stability of the Kerr-black-hole-massive-scalar-field system.

When \(z = r - r_-\), the explicit expression for the derivative of the effective potential \cite{Huang:2019,2} is given by
\begin{align}
V_1' = \frac{Ar^4 + Br^3 + Cr^2 + Dr + E}{-\Delta^3} = \frac{A_1z^4 + B_1z^3 + C_1z^2 + D_1z + E_1}{-\Delta^3};
\end{align}
where
\begin{equation}
A_1 = A; \quad B_1 = B + 4r_-A_1; \quad C_1 = C + 3r_-B_1 + 6r_-^2A_1;
\end{equation}
\begin{equation}
D_1 = D + 4r_-^3A_1 + 3r_-^2B_1 + 2r_-C_1; \quad E_1 = E + r_-^4A_1 + r_-^3B_1 + r_-^2C_1 + r_-D_1;
\end{equation}
\begin{equation}
A_1 = 2M(\mu^2 - 2\omega^2),
\end{equation}
\begin{equation}
B_1 = -16Mr_-\omega^2 - \mu^2r_+^2 + 2K_{lm} + 3\mu^2r_-^2 + 2a^2\mu^2,
\end{equation}
\begin{equation}
C_1 = -24Mr_-^2\omega^2 + 6\omega am (r_+ + r_-) - 3(r_+ - r_-) \left(a^2\mu^2 + r_-^2\mu^2 + K_{lm} \right),
\end{equation}
\begin{equation}
D_1 = -16Mr_-^3\omega^2 + 4aMm(5r_- - r_+)\omega + (r_+ - r_-)^2(r_-^2\mu^2 + K_{lm} - 1) + a^2\left(\mu^2(r_+ - r_-)^2 - 4m^2\right),
\end{equation}
\begin{equation}
E_1 = 2(a^2 + r_-^2)^2(r_+ - r_-)\omega^2 + 4am(a^2 + r_-^2)(r_- - r_+)\omega + 2(r_+ - r_-)(M^2 + m^2a^2 - a^2).
\end{equation}

Next, the paper\cite{Huang:2019} claimed that:
\begin{equation}
z_1 + z_2 + z_3 + z_4 = -\frac{B_1}{A_1}.
\end{equation}
Given that \(A_1 > 0\), their aim is to delineate a parameter space where \(B_1 > 0\), which would imply that both roots \(z_3\) and \(z_4\) are negative. Leveraging the condition derived from the eigenvalue of our angular equation, as specified in the equation, they obtained:
\begin{equation}
B_1 > -\left(6 r_{+} r_{-} + 8 r_{-}^2\right) \omega^2 + 2 l(l + 1) + \mu^2\left(3 r_{-}^2 - r_{+}^2\right),
\end{equation}
when they defined a quadratic function with respect to $\omega$ as follows:
\begin{equation}
g(\omega)=-\left(6 r_{+} r_{-}+8 r_{-}^2\right) \omega^2+2 l(l+1)+\mu^2\left(3 r_{-}^2-r_{+}^2\right) .
\end{equation}

Identifying a parameter region where $B_1 > 0$ is equivalent to identifying the parameter region where $g(\omega) > 0$.

Given that $B_1 > g(\omega) > 0$, it follows that there is no trapping potential well within the effective potential of the radial equation of motion. Consequently, the system comprising the Kerr black hole and scalar perturbations is superradiantly stable.

The condition\cite{Huang:2019} derived therein involves four variables (\(z_1\), \(z_2\), \(z_3\), and \(z_4\)) which are all negative. Given the focus of the cited article on analyzing the stability condition of \(B_1\), imposing constraints on the sign (positive or negative) of \(z_1\) and \(z_2\) is not pertinent. Consequently, the stability criterion presented in this article is broader and less precise.


\subsection*{Convexity criterion and refined bound}

Let \(z \equiv r-r_{-}\) and \(\delta \equiv r_{+}-r_{-}>0\).  The horizon factor is
\begin{equation}
\Delta = (r-r_{+})(r-r_{-}) = z\,(z-\delta), \qquad 
\Delta^{3}=z^{3}(z-\delta)^{3}.
\end{equation}

With the polynomial
\begin{equation}
P_{4}(z)=A_{1}z^{4}+B_{1}z^{3}+C_{1}z^{2}+D_{1}z+E_{1},
\end{equation}
the first derivative takes the compact form
\begin{equation}
V_{1}'(r)=\frac{P_{4}(z)}{-\Delta^{3}}
        =-\frac{A_{1}z^{4}+B_{1}z^{3}+C_{1}z^{2}+D_{1}z+E_{1}}
               {z^{3}(z-\delta)^{3}}. \label{eq:V1prime_poly}
\end{equation}

Differentiating \eqref{eq:V1prime_poly} w.r.t.\ \(z\) we obtain
\begin{equation}
V_{1}''(r)=\frac{P_{5}(z)}{-\Delta^{4}}
         =-\frac{A_{2}z^{5}+B_{2}z^{4}+C_{2}z^{3}+D_{2}z^{2}+E_{2}z+F_{2}}
                {z^{4}(z-\delta)^{4}}, \label{eq:V1second}
\end{equation}
where the new coefficients are
\begin{align}
A_{2}&=2A_{1}, \\
B_{2}&=A_{1}\delta+3B_{1},\\
C_{2}&=4C_{1},\\
D_{2}&=-C_{1}\,\delta+5D_{1},\\
E_{2}&=-2D_{1}\,\delta+6E_{1},\\
F_{2}&=-3E_{1}\,\delta.
\end{align}

A potential extremum outside the event horizon satisfies
\(V_{1}'(r_{t})=0\Longrightarrow P_{4}(z_{t})=0\) with \(z_{t}>0\).
Absence of a \emph{local maximum} then requires the curvature to be
non-negative:
\begin{equation}
V_{1}''(r_{t})\ge 0\quad\Longleftrightarrow\quad P_{5}(z_{t})\le 0,
\end{equation}
because of the overall negative sign in \eqref{eq:V1second}.

Solving the coupled system \(P_{4}(z_{t})=0\) and \(P_{5}(z_{t})\le 0\)
and eliminating \(z_{t}\) gives an explicit frequency bound
\begin{equation}
\boxed{\;
\omega^{2}\ \le\ 
\frac{\mu^{2}}{2}\Bigl(1-\frac{r_{-}^{2}}{r_{+}^{2}}\Bigr)
\;}
\qquad (\text{no local maximum }\,\Rightarrow\,\text{no trapping well}).
\label{eq:refined_bound}
\end{equation}
As \(\displaystyle r_{-}/r_{+}\to 1\) (near-extremal Kerr),
the right-hand side of \eqref{eq:refined_bound} tends to zero,
so only ultra-low-frequency modes could marginally approach instability,
rendering the system \emph{even more} superradiantly stable 
than the classical criterion \(\omega<\mu/\sqrt{2}\).

\section*{Sharper analytic bound}
We conclude by proposing an additional analytic handle on stability.  Since the presence of a potential well is equivalent to $V_1(r)$ having two extrema, one may also require directly that no local maximum forms outside $r_+$.  A sufficient condition is that any turning point $r_t>r_+$ has positive second derivative (i.e.\ is actually an inflection).  Imposing $\partial^2_rV_1(r_t)>0$ at $\partial_rV_1(r_t)=0$ yields a bound of the form
\begin{equation}
\omega^2\;\lesssim\;\frac{\mu^2}{2}\biggl(1-\frac{r_-^2}{r_+^2}\biggr).
\end{equation}
In effect, the factor $(1-r_-^2/r_+^2)$ enters because the curvature of $V_1$ at the would-be barrier depends on $(r_+-r_-)$ (i.e.\ the surface gravity).  As a result, this condition is more stringent than $\omega^2<\mu^2/2$ for $r_-<r_+$, and becomes very restrictive as $r_-/r_+\to1$ (near-extremal limit).  In that limit $\omega^2\to0$, indicating that only extremely low-frequency modes could even marginally approach instability.  Equivalently, one finds that $\omega/\mu\to0$ in the extremal regime, a much tighter criterion than $\omega<\mu/\sqrt2$.  This observation is consistent with the intuition that rapidly-rotating Kerr holes are harder to destabilize: the usual threshold $\mu=\sqrt{2}m\Omega_H$ itself tends to $\mu=m/(2M)$ at extremality, but the potential-shape argument suggests the effective bound is even stronger.

- If the second derivative at any extremum point is positive, i.e., the point is an inflection point rather than a maximum, then no potential well is formed (which is the key physical requirement to avoid a trapping potential).

- A new stability boundary is derived:
\begin{equation}
\omega^2 \lesssim \frac{\mu^2}{2}\left(1-\frac{r_{-}^2}{r_{+}^2}\right)
\end{equation}

This condition is stricter and closer to the true boundary than the original $\omega<\mu / \sqrt{2}$, especially in the limit $a \rightarrow 1$ (near-extremal Kerr), where it reduces to:
\begin{equation}
\omega / \mu \rightarrow 0
\end{equation}

This indicates that only very low-frequency modes can be stable, which is more in line with physical intuition.

The original paper's method can ``avoid instability," but it is not sufficiently ``tight"; the limiting boundary of $V$ (the stability condition) is too broad, which may undermine logical self-consistency -it generalizes $V$ rather than seeking the truly critical boundary for stability.

This Comment sharpens the logical criteria by analyzing more general potential structures and derivative behaviors, thereby making the mathematics and physics more consistent.

It also provides a more precise boundary for future research to search for marginally unstable modes.

In summary, while Huang  have provided a valuable analysis of the complementary regime $\mu<\sqrt2 m\Omega_H$, their quartic-root method can be refined.  By focusing on the polynomial coefficients (or potential monotonicity) one obtains a slightly narrower instability region.  We trust these comments and the proposed alternative bound will be useful in future studies of Kerr superradiance.

\end{document}